# Monitoring ion track formation using *in situ* RBS/c and ERDA

M. Karlušić[1], S. Fazinić[1], Z. Siketić[1], T. Tadić[1], D. Ćosić[1], M. Jakšić[1], M. Schleberger[2]

[1]*Ruđer Bošković Institute, Bijenička cesta 54, 10000 Zagreb, Croatia*

[2]*Fakultät für Physik and CENIDE, Universität Duisburg-Essen, D-47048 Duisburg, Germany*

[*] Corresponding author: marko.karlusic@irb.hr

## Abstract

The aim of this work is to investigate feasibility of the ion beam analysis techniques for monitoring swift heavy ion track formation. First, use of the *in situ* Rutherford backscattering spectroscopy in channeling mode to observe damage build-up in quartz $SiO_2$ after MeV heavy ion irradiation is demonstrated. Second, new results of the *in situ* grazing incidence time-of-flight elastic recoil detection analysis used for monitoring the surface elemental composition during ion tracks formation in various materials are presented. Ion tracks were found on $SrTiO_3$, quartz $SiO_2$, a-$SiO_2$ and muscovite mica surfaces by atomic force microscopy, but in contrast to our previous studies on GaN and $TiO_2$, surface stoichiometry remained unchanged.

1. ## Introduction

Passage of swift heavy ions (SHI) through the materials may result in the formation of permanent damage along their trajectories called ion tracks. These nanoscopic objects are formed by a number of different physical processes that span femtosecond to nanosecond time domains. Ion tracks have been investigated for many years, and several review papers documents well the development of this research field [NI09], [FA11], [MT12], [FAL16]. Still, many questions related to basic understanding of ion track formation are open and subject to vigorous scientific investigations [MT12], [SK06], [GS11], [MT12b], [MK12], [GS13], [MK17]. Besides basic research, clarifications to these hotly debated issues could have implications for ion track applications as well [MT04], [MT09].

Historically, ion track research has been spearheaded by research groups at large accelerator facilities. Reason for this is related to the existence of a threshold for ion track formation, i.e. ion tracks can be formed only if density of deposited energy exceeds certain critical value. According to thermal spike models, phase transition (most often melting of the material) is necessary requirement for ion track formation that has to be triggered by sufficiently high density of deposited energy. Only then, relaxation of the deposited energy results in formation of permanent damage, which is otherwise dissipated away without noticeable damage to the material. Various materials have widely different thresholds for ion track formation, and quite often high threshold values require SHI with kinetic energies in the 100-1000 MeV range for systematic investigations. These energies are nowadays accessible worldwide, but noticeably ion track research now also takes place at medium accelerator facilities where kinetic energies in the range of ~10 MeV are available. At these energies, electronic stopping power is still dominant mechanism of the SHI kinetic energy deposition. Furthermore, capability for ion track production given by the electronic stopping power of the SHI is highly nonlinear function of its kinetic energy, and therefore medium accelerator facilities can deliver SHI beams of interest for ion track studies in many materials [MB12], [IBR12], [MK13], [IBR16]. Often it is possible to pinpoint exactly the threshold for ion track formation. Practical considerations like limited availability of the beamtime can also play a significant role at large facilities, thus it is advantageous for complementary investigations at lower energies to be outsourced elsewhere, as our own research documents well [MK10], [MK15], [MK15b], [MK16], [MK16b], [MK17], [MK17b]. SHI applications can benefit from this also, for example hadron therapy using carbon ions typically requires ~100 MeV beams in order to reach targeted volume. But electronic stopping power maximum for carbon ions in water is at 5 MeV, thus basic research related to understanding hadron therapy can be easily done at much lower energies.

In this work, we aim to present another important aspect of ion track studies at medium accelerator facilities, namely access to ion beam analysis (IBA) techniques that are practically not available at large accelerator facilities. IBA techniques are set of powerful and versatile material science techniques that can provide elemental depth profiles, and by use of ion microprobes also high resolution elemental or density distribution maps can be acquired. For example, we have recently demonstrated imaging of etched ion tracks by scanning transmission ion microscopy (STIM) [MK13] and biological microstructures by MeV secondary ion mass spectrometry (MeV SIMS) [ZS15]. Furthermore, ion microprobes can be used as a tool for nanoscale material patterning, like production of ordered arrays of etched single ion tracks [RWS12] and MeV ion lithography [MV11].

Few of the IBA techniques, like Rutherford Backscattering in channeling mode (RBS/c), ion beam induced charge (IBIC) and ionoluminescence (IL) can also provide information about defects in monocrystalline samples. Actually, RBS/c is one of the most often used techniques for ion track measurements because it is very suitable for monitoring damage build-up during SHI irradiation. While transmission electron microscopy (TEM) and atomic force microscopy (AFM) enable direct observation of ion tracks, RBS/c has been used in many studies for detailed studies of ion track evolution vs. electronic stopping power [AM94], [MT12], [MT12b], [OPR12]. Damage kinetics observed by RBS/c provides additional information about mechanism of ion track formation, i.e. weather multiple SHI hit is necessary for amorphization of the material or is it a single SHI process [SMMR98], [MT01], [AR10], [OPR12]. Detailed monitoring of damage build up within the material during SHI irradiation is in particular of interest close to the ion track formation threshold, where deviations from simple overlap track models are expected [SMMR98]. However, such detailed studies require irradiation and RBS/c measurement on large number of samples, and these experiments pose large demand on the beamtime. Recently, solution to this problem has been found in the establishment of experimental set-ups for *in situ* measurements using different analytical techniques like IL [EG13], [NM15], [MLC16], X-ray diffraction [CG12], AFM [FM16] and Raman Spectroscopy [SD15], [SM15].

Here we present two approaches for *in situ* measurements of the ion tracks based on the RBS/c and elastic recoil detection analysis (ERDA). In the first case, we describe dual-beam chamber where materials modification can be induced by SHI delivered from the 6 MV EN Tandem Van de Graaff accelerator, while monitoring of the damage kinetics can be done simultaneously by RBS/c using probing beam from the 1 MV Tandetron accelerator. In the second example we show how ERDA can be utilized for monitoring the surface stoichiometry during the grazing incidence SHI irradiation that results in very long ion tracks on the material surface. Previously we have shown that ion track formation in GaN [MK15] and $TiO_2$ [MK16] is accompanied by the loss of nitrogen and oxygen, respectively. However, ion track formation in $CaF_2$ does not result in changes of the surface stoichiometry under the same irradiation conditions [MK17]. Here we present results of similar ERDA measurements made on another track forming materials, namely quartz $SiO_2$, amorphous $SiO_2$, $SrTiO_3$ and muscovite mica.

## 2. Experiment

### 2.1. Monitoring ion track production in quartz SiO$_2$ by *in situ* RBS/c

Epi-ready single crystal quartz SiO$_2$[0001] samples were purchased from Crystec (Germany). All ion irradiations were performed at the Ruđer Bošković Institute accelerator facility [MJ07]. For the present study, the experiment was performed in the dual beam chamber equipped with 6-axis goniometer (Figure 1a) using 5 MeV Si, 4 MeV C, 2 MeV Li and 1 MeV p beams delivered from the 1 MV Tandetron accelerator. For the RBS/c measurement, probing ion beam had 1 mm beam spot in diameter, and the current was kept at around 1 nA. To detect backscattered ions, a silicon surface barrier (SSB) detector was positioned at 160° with respect to the probing beam direction. The angular scan maps (tilt, azimuth) were acquired for target alignment.

For *in situ* RBS/c analysis it is important that repeated RBS/c measurements at the same spot do not introduce additional damage to the material under investigation. Therefore, we have performed test measurements using 2 MeV Li beam on the ion radiation sensitive CaF$_2$ crystal (Korth Kristalle, Germany) containing defects introduced previously by 23 MeV I [MK17]. This way it was verified that the RBS/c beam does not introduce additional defects after prolonged exposure, and that RBS/c spectra from irradiated samples (containing disorder) can be reliably acquired even after multiple probing beam exposures, as shown on Fig. 1(b).

While more energetic ion beams from 6 MV EN Tandem Van de Graaff accelerator can be inserted into the dual beam chamber as well, for the present study of damage kinetics in quartz SiO$_2$, 5 MeV Si and 4 MeV C beams delivered by 1 MV Tandetron accelerator were found sufficient. Irradiations were done on two 1×1 cm$^2$ single crystal quartz SiO$_2$[0001] samples that were irradiated on several positions by different fluences of 5 MeV Si and 4 MeV C beams, having electronic stopping powers $S_e$ = 3.2 keV/nm and $S_e$ = 1.55 keV/nm, respectively [JFZ10]. For these two ion beams, nuclear stopping powers were $S_n$ = 0.03 keV/nm and $S_n$ = 0.003 keV/nm, respectively [JFZ10]. Irradiations by 5 MeV Si beam were done at 6° off normal, while irradiations by 4 MeV C beam were done both in the channeling (on-axis) and 6° off normal (i.e. random). RBS/c probing beam in both cases was 1 MeV p.

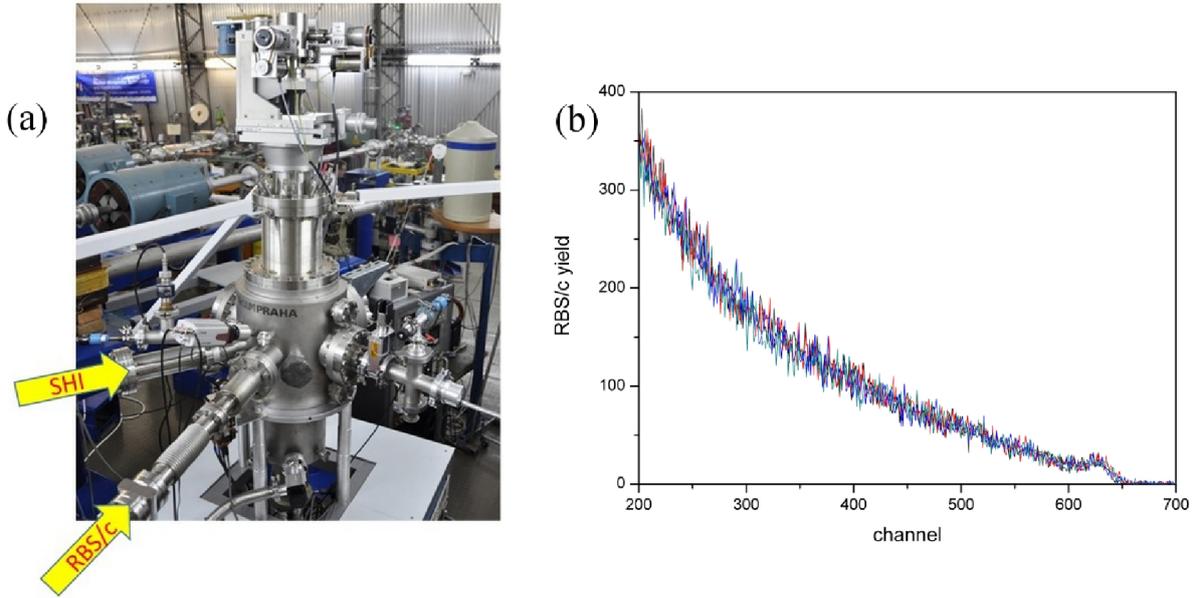

**Figure 1.** (a) Dual beam vacuum chamber for simultaneous SHI irradiation and *in situ* RBS/c measurements. (b) Four successive RBS/c spectra obtained from an irradiated $CaF_2$ sample (23 MeV I, fluence $3\times10^{12}$ cm$^{-2}$).

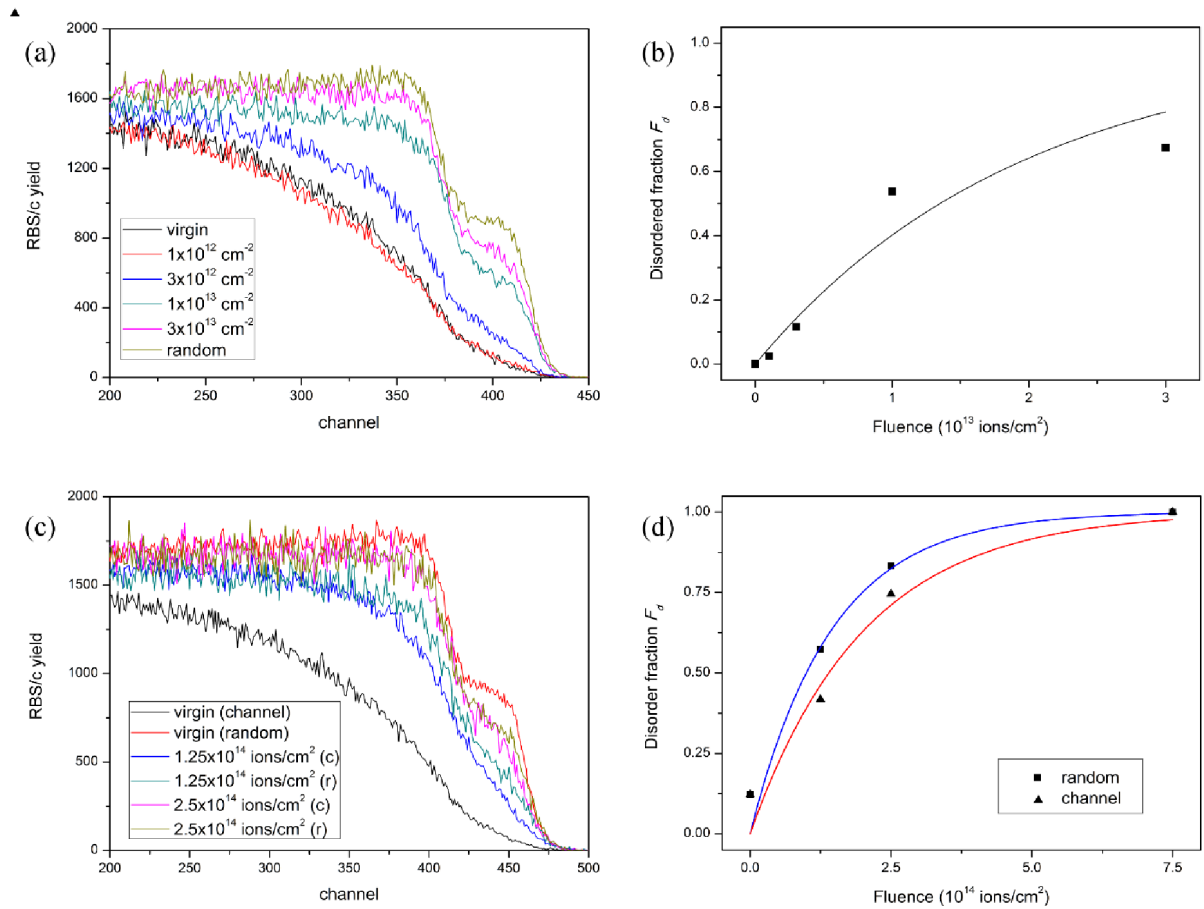

**Figure 2.** (a) RBS/c spectra from quartz $SiO_2$ irradiated with different fluences of 5 MeV Si beam. (b) Analysis of RBS/c data yields ion track radius $R = 1.3\pm0.15$ nm. (c) RBS/c spectra from quartz $SiO_2$ irradiated with different fluences of 4 MeV C beam, both in channeling and random conditions. (d) Analysis of RBS/c data yields damage cross sections $\sigma_R = 0.7\pm0.1$ nm$^2$ (blue line) and $\sigma_C = 0.5\pm0.1$ nm$^2$ (red line) for the random and channeling irradiations, respectively.

For the analysis of the 5 MeV Si irradiated sample, the highest energy part of RBS/c spectra (Fig. 2a) coming from backscattered protons on the Si sublattice was used to calculate amount of disorder, i.e. disordered fraction $F_d$. Using known procedure based on the surface approximation, the ion track radius $R$ can be derived from the Poisson's law that describes the evolution of the disordered fraction $F_d$ with the applied SHI fluence $\Phi$ [AM94], [MT12b], [MK17]. As shown in Fig. 2b, RBS/c spectra provide evidence that gradual disordering of the quartz $SiO_2$ sample takes place with increasing 5 MeV Si fluence, and for the highest applied fluence, almost complete amorphization takes place. From the observed damage kinetics, we evaluated ion track radius $R = 1.3\pm0.15$ nm, in agreement with previous works [AM94], [OPR12].

Analysis of the RBS/c spectra from the 4 MeV C irradiated sample follows a slightly different procedure [GG15], and spectrum integration between channels 425-475 (Fig. 2c) is taken as a measure of the disorder. This approach was found adequate for the analysis of damage occurring below threshold for ion track formation, and electronic stopping power of 4 MeV C beam is below threshold of 2 keV/nm [OPR12]. While the process causing damage is not clear (nuclear stopping power or self-trapped exciton mechanism), we show in Figs. 2c,d that presented *in situ* RBS/c set-up will enable detailed measurements of the channeling and near-channeling effects influencing ion track formation [MT09], [TS08], [MK17b].

### 2.2. Evaluating ion track stoichiometry by *in situ* ERDA

Epi-ready quartz $SiO_2$ and $SrTiO_3$ single crystal samples were purchased from the Crystec (Germany). Thin amorphous $SiO_2$ film (200 nm) grown on Si wafer purchased from Crystec was also found suitable for the AFM analysis. Muscovite mica samples were obtained from 2SPI, and surfaces were freshly cleaved prior to irradiation.

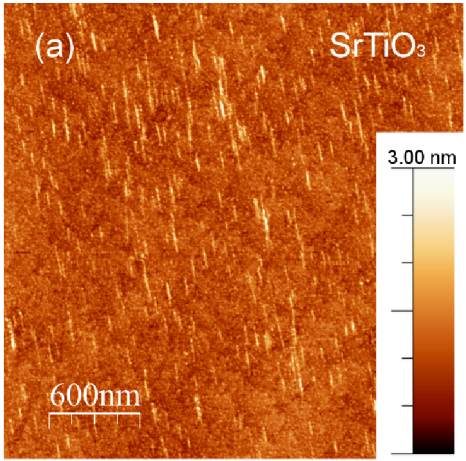
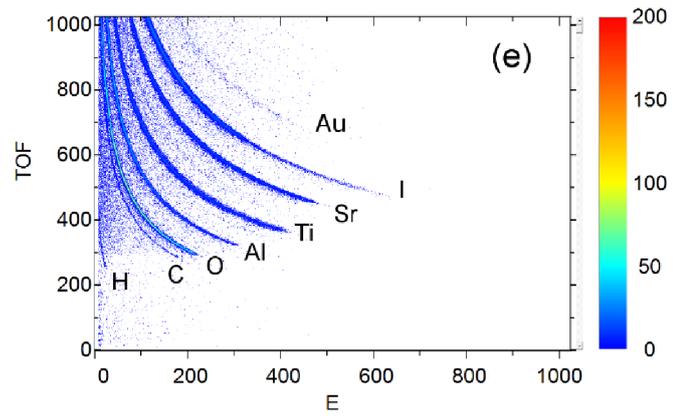
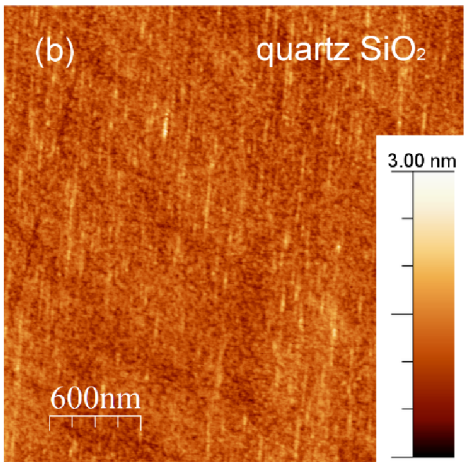
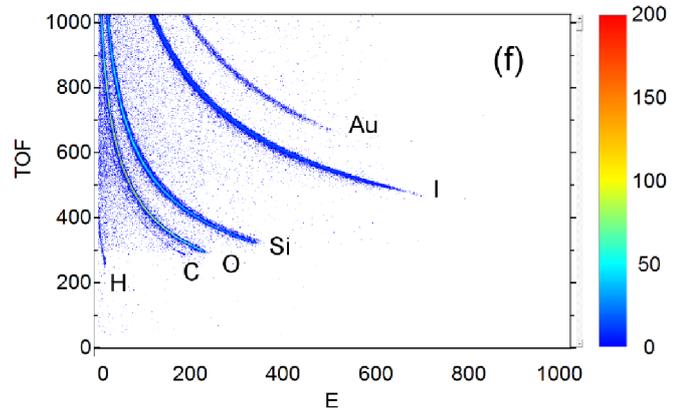
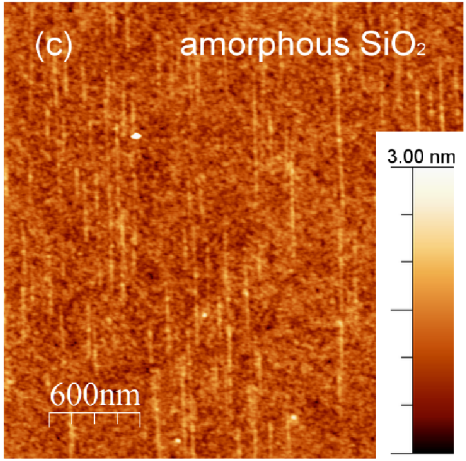
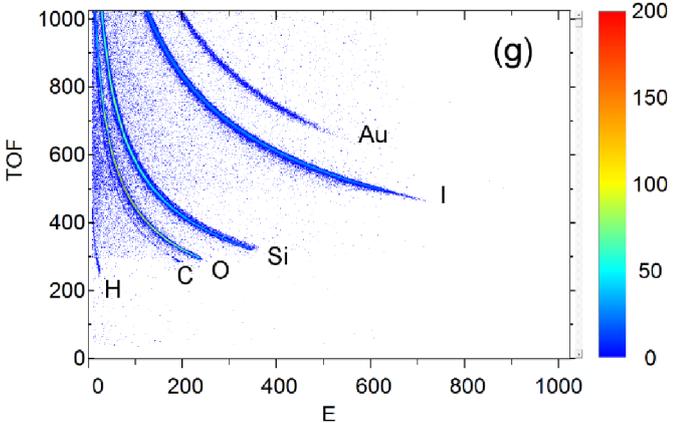
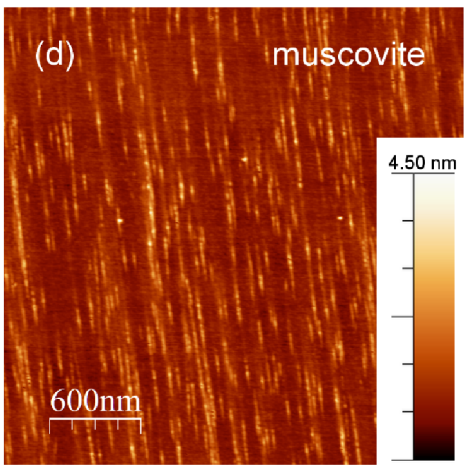
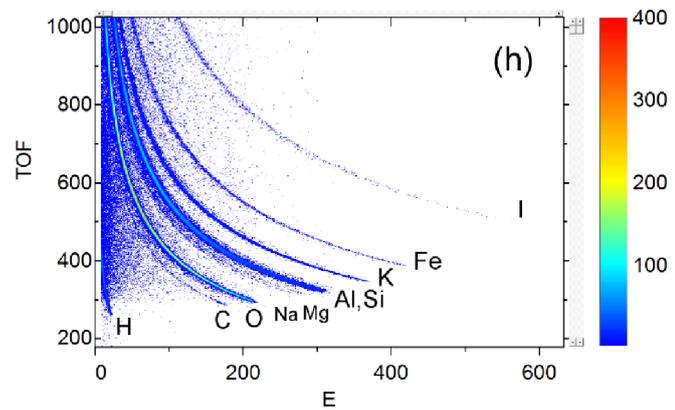

**Figure 3.** Ion tracks on the materials surfaces produced by 23 MeV I under the grazing incidence angle of 1°, observed by AFM: (a) SrTiO$_3$, (b) quartz SiO$_2$, (c) amorphous SiO$_2$, (d) muscovite mica. In all the cases, applied ion fluence matches well the observed ion track density. Corresponding grazing incidence ERDA spectra obtained by the same 23 MeV I beam are shown in panels (e)-(h). Detection of iodine is from the primary ion beam, while detected Al and Au are from the gold coated aluminium sample holder.

For the ERDA study, grazing incidence irradiation was done using 23 MeV I beam delivered by the 6 MV EN Tandem Van de Graaff accelerator, and time of flight (ToF) ERDA was carried out at the dedicated chamber [ZS08], [ZS10]. The ToF-ERDA measurements were performed at 1° grazing incidence angle with respect to the sample surface, with spectrometer positioned at the angle of 37.5° towards the beam direction. All the data were collected in the 'list mode' and offline replay/analysis with sections was performed using the Potku software package [KA14]. Afterwards, ion tracks on the surfaces were analysed using tapping mode AFM. The AFM measurements were performed under ambient conditions using a Dimension 3100 AFM and Nanosensors NCHR cantilevers at Universität Duisburg-Essen. Images were analysed using the WSxM code [IH07].

Ion tracks were formed on surfaces of all irradiated materials after exposure with the 23 MeV I beam under 1° grazing incidence angle. Observed surface tracks consist of well-known intermittent structure that is easily recognizable by AFM, as Fig. 3 shows. For all investigated samples, density of the ion tracks agrees well with the applied ion fluence, i.e. efficiency of the ion track formation is close to 1. Fig. 3 also shows ToF-ERDA spectra collected during exposure to the same SHI beam and the same 1° grazing incidence angle. Although irradiation angle of 1° is not common for ToF-ERDA, it was still possible to measure stoichiometry on and below the material surface with depth resolution of 5 nm. For all investigated materials, the surface elemental composition remained unchanged. Therefore, we conclude that ion track formation in investigated materials is not accompanied by the preferential loss of any element.

## 3. Discussion

Recently, there is an increase in research interest for *in situ* ion track measurements using IBA techniques [WMA05], [JJ10], [CG12], [EG13], [MK15], [FM16], [MK16], [MK17]. Numerous data points that can be acquired on a single sample during the single irradiation run can save a lot of valuable beamtime and provide large amount of data necessary for detailed investigation of damage kinetics [SMMR98], [MT01], [AR10], [OPR12]. Besides our on-going efforts to utilize grazing incidence ToF-ERDA for *in situ* elemental analysis of ion tracks, here we introduce *in situ* RBS/c for monitoring damage build-up due to the production of ion tracks. This is especially important for ion track studies done close to the threshold for ion track formation. Below the threshold, subthreshold damage arising from the secondary damage mechanism (for example exciton mechanism in the case of $TiO_2$ [AR10]) can be identified by deviation of the experimental data from Poisson law. Similarly, above the threshold behaviour described by Avrami equation can also indicate discontinuities within small ion tracks [SMMR98]. Experimental data we show on Fig. 2 are consistent with the results published before [OPR12], but deviation from Poisson law due to the proximity of the ion track formation threshold can be suspected. While more detailed damage kinetics is probably not possible to achieve using proton beams, use of the 2 MeV Li beams requires significantly longer time for the collection of the RBS/c spectra. Therefore, measurements using He beams (to be available after the upgrade of the ion source system) remain a future work to be done in our laboratory.

Although IBA techniques are generally considered to be non-destructive, clearly one has to be careful about the possible damage introduced by the probing RBS/c beam. Care has to be taken particularly for He and Li ion beams that have much larger stopping powers than proton beams. Surprisingly, we have found that even for sensitive material like $CaF_2$, the damage of the probing 2 MeV Li beam is below the detection limit, and RBS/c measurements repeated on the same spot on the sample surface yield the same result. This is in stark contrast with great sensitivity of this material to the e-beam induced damage and related difficulties during the TEM observations [JJ98], [JJ98b], [MK17]. Although the RBS/c probing beam dissipates energy within $CaF_2$ via electronic excitations, secondary electrons generated this way have much lower energy than e-beam used in the TEM. Therefore, damage to the material via electronic excitations is avoided and use of RBS/c data is clearly better choice for thermal spike studies than TEM data [MK17].

Another important aspect of the *in situ* measurements can be related to specific conditions when samples have to be investigated without breaking the vacuum. For example, coverage by water layer was observed during AFM studies of the ion tracks on the $CaF_2$ surfaces [NK06], [EG16] and clean Si surfaces are prone to very fast oxidation [OO12], [OO13]. Furthermore, 2D materials like graphene and $MoS_2$ can become chemically reactive by the ion introduced defects, especially on the place of the ion impact or on the edge of the ion-introduced pores [MK15b], [MK16b], [LM17]. By taking the irradiated samples out of the vacuum, follow-up measurements under the ambient conditions can be affected and for this reason AFM measurements should be preferably done in vacuum [FM16]. Since Raman spectroscopy is very powerful analytical technique that directly can probe into damage kinetics of the 2D materials [MK15b], [MK16b], [JZ16], *in situ* Raman spectroscopy set-ups that have been commissioned over the last few years [SD15], [SM15] should be very interesting for ion-irradiation studies of 2D materials.

Regarding *in situ* ERDA measurements, these have been used before to monitor stoichiometry changes during the SHI irradiation [DKA01], [WMA05], [JJ10]. Results presented here and in our previous publications [MK15], [MK16], [MK17] demonstrate *in situ* ToF-ERDA performed under the 1° grazing incidence angle, thus enabling monitoring of the stoichiometry changes under conditions for surface ion track formation. With the exception of GaN that obviously decomposes during the thermal spike produced by the ion impact [MK15], and still unclear mechanism of preferential oxygen loss from $TiO_2$ [MK16], all other investigated materials show constant stoichiometry during SHI irradiation.

**Conclusion**

For the first time we have successfully demonstrated *in situ* RBS/c measurements of the ion tracks. Here presented dual beam set-up will enable precise monitoring of the damage build-up within the material during SHI irradiation. Surprisingly, RBS/c using heavier ions like lithium instead of protons, does not introduce additional damage to sensitive materials like $CaF_2$. Therefore, multiple RBS/c measurements on the same spot yield reliable results that are necessary for *in situ* RBS/c. This approach is especially important for studies close to the ion track formation threshold, where deviations from simple overlap track models are expected.

To accomplish this, a substantial amount of experimental data is needed for reliable analysis, and this novel approach could provide adequate solution to this challenge.

We also present new experimental data on the *in situ* grazing incidence ToF-ERDA measurements. This way, elemental composition of the surface can be monitored under the conditions when surface ion tracks are produced. Any change in the stoichiometry of the surface provides evidence about elemental composition of the ion tracks. In the present study, we find that surface stoichiometry of the investigated materials ($SrTiO_3$, muscovite mica, quartz $SiO_2$, and a-$SiO_2$) remains unchanged.

## Acknowledgements

MK, SF, ZS, TT and MJ acknowledge the financial support from the Croatian Science Foundation (pr. No. 8127). Support from the Croatian Centre of Excellence for Advanced Materials and Sensors is also acknowledged. The authors acknowledge the CERIC-ERIC Consortium for the access to experimental facilities.